# Generative AI May Prefer to Present National-level Characteristics of Cities Based on Stereotypical Geographic Impressions at the Continental Level

Shan Ye[1]

[1]School of Information Engineering, China University of Geosciences (Beijing)

**Abstract**

A simple experiment was conducted to test the ability of the Chinese-based generative artificial intelligence platform, Wenxin Yige, to render urban street views of different countries. The study found that the cityscape images generated by this AI platform may contain continental-level stereotypes in terms of showing the level of economic development and modernization. Street view images generated from Wenxin Yige do not adequately represent the diverse range of urban landscapes found across different nations. Using these generated images for geography education or outreach initiatives could inadvertently strengthen people's existing stereotypical views about individual countries.

**Introduction**

Presently, large language models and associated generative artificial intelligence (AI) technologies are emerging as pivotal trends in the realm of AI research and development. The capabilities demonstrated by these generative AI systems have been both impressive and promising, as evidenced by the growing popularity of AI-powered chatbots and text-to-image systems. This advancement holds new potential across a myriad of application sectors such as business operations, management strategies, academic research, and scientific outreaches (e.g. Castelli and Manzoni, 2022; Weisz et al., 2023; Gozalo-Brizuela and Garrido-Merchán, 2023). Given the rapid development of generative AI today, many researchers have turned their attention to how well generative AI grasps geographical knowledge. For instance, Roberts et al. (2023) conducted a series of experiments, ranging from basic geography Q&A to outlining countries and continents as well as route planning, in order to study how the large language model of GPT-4 understands world geography - they found that GPT-4 generally has a good understanding of it. Jang et al., (2023) analyzed how generative AI handles place identity and discovered that both GPT-4 and DALL-E2, a text-to-image generator, can distinctly differentiate between cities around the world.

However, as generative AI gains traction, concerns regarding its accuracy and accompanying ethical implications have concurrently surfaced. For instance, Srinivasan and Uchino (2021) explored biases in AI-generated art through the lens of art history. Their findings revealed that AI-created artworks, particularly those involving style transfer techniques, are susceptible to various forms of bias including dataset bias and selection bias among others. These biases can exert significant social and cultural influence (Srinivasan and Uchino, 2021). In a separate study by Smith and Williams (2021), it was found that inherent biases within training data could result in chatbots failing to adequately address issues related to gender, race, and occupation equitably. This shortcoming may culminate in the manifestation of stereotypes within generated dialogues. Furthermore, Park (2023) highlighted potential adverse outcomes resulting from biased generative AI models such as influencing human decision-making processes or potentially propagating misinformation via generated content.

Potential adverse effects of generative AI on geographic education and related knowledge dissemination also warrant attention. For instance, Kang et al. (2023) conducted pioneering research into issues associated with maps generated by the DALL-E2 model. Their findings indicated that these generated maps could contain inaccuracies, misleading features, and potentially instigate commercial and copyright disputes; all of which pose significant challenges if such maps are employed in educational activities. Scheider et al. (2023) underscored that responses from generative AI chatbots like ChatGPT currently lack the robustness required for effective teaching in geography and GISciences. Beyond concerns about accuracy, it is imperative to consider whether generative AI models harbor other biases or stereotypes specifically pertaining to geography - if they do exist, understanding their extent becomes a topic of considerable interest. In this context, the generative AI system known as Wenxin Yige (hereafter referred to as Wenxin) is utilized as an exemplar to illustrate potential geography-related stereotypes present within text-to-image generative AIs through a basic analysis.

**Data and Method**

Released in 2022, Wenxin is an AI-powered arts platform provided by Baidu and is one of the popular and accessible text-to-image generators in China (https://yige.baidu.com/). The platform can generate images based on inputs of users in simple Chinese natural languages. The purpose of this experiment is to examine how the AI behind the Wenxin platform understands cities in different countries and if there are any underlying stereotypes.

In order to quantify the results, we bifurcate the Wenxin's impressions of cities into two distinct dimensions: developed-underdeveloped and traditional-modern. The developed-underdeveloped dimension seeks to ascertain whether the Wenxin model is capable of discerning differences in economic development levels between cities across various countries, and if so, does this distinction correspond with reality. On the other hand, the traditional-modern dimension aims to determine whether city images generated by Wenxin for different countries predominantly feature traditional architectural elements and landscapes or lean towards modernization.

The initial step involves discerning how the Wenxin model interprets the aforementioned aspects of cityscapes in the generated results. Given that outputs from generative AI may not perfectly align with reality, it would be imprudent to utilize real-world images for this test. Consequently, as a preliminary phase of this experiment, simple and unambiguous natural language is input into the Wenxin system to generate street view images of unspecified cities. In the first set, 500 street view images are generated under the descriptor "economically underdeveloped cities", followed by another 500 designated as "economically developed cities". These two sets will serve in evaluating how the Wenxin system perceives the economic development spectrum within urban landscapes in the next step. Similarly, 500 images each for "traditional cities" and "modern cities" are generated to assess how Wenxin interprets cityscapes along the traditional-modern axis.

Examples of generated images from each set are presented in Figure 1. It is evident that the Wenxin model distinctly differentiates between cities based on their economic development status and the preservation of traditional elements. In the images produced by Wenxin, streetscapes of economically underdeveloped cities typically exhibit a certain level of disarray with older, run-down buildings (Figure 1a). Conversely, those belonging to developed cities manifest traits of neatness and orderliness (Figure 1b). Regarding the other dimension, imagery associated with traditional cities as rendered by Wenxin predominantly feature low-rise structures, often bordered by narrow cobblestone streets and interspersed with occasional traditional embellishments such as lanterns adorning buildings (Figure 1c). Contrarily, generated street view depictions of modernized cities frequently showcase high-rise residential or commercial skyscrapers (Figure 1d). To human observers, these characteristics are readily distinguishable.

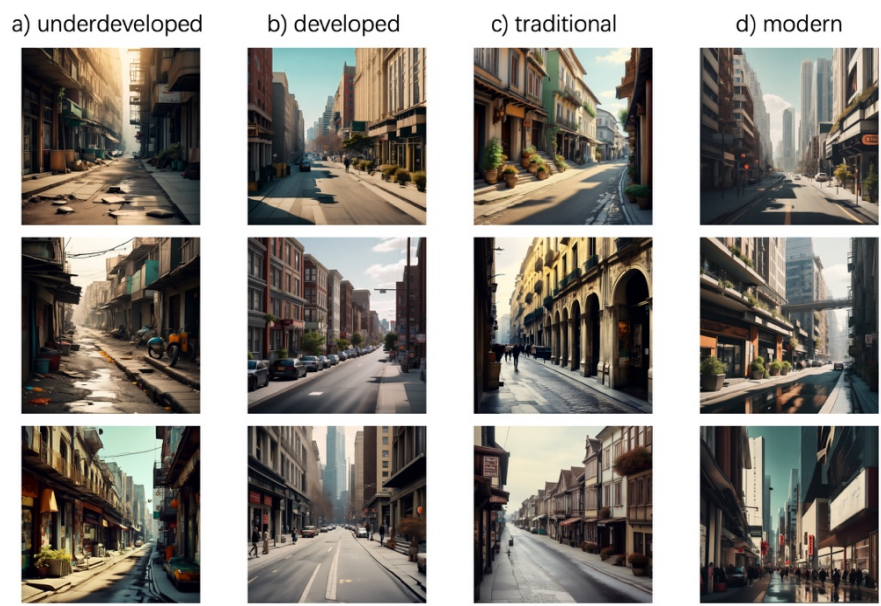

*Figure 1: Examples of generated figures from the Wenxin platform in four categories: a) economically underdeveloped, b) economically developed; c) traditional; and d) modern.*

Subsequently, two deep learning-based detectors are developed to quantify the capacity of the Wenxin model to distinguish between these two dimensions. In this context, the ResNet model is selected since it is a classic and widely utilized deep convolutional neural network known for its efficacy in image classification (He et al., 2016). The first detector is designed to discern whether a generated image represents an economically developed city. The 1000 images from 2 sets pertaining to the underdeveloped-developed dimension are partitioned into training, validation, and testing datasets at a ratio of 7:2:1. The ResNet50 model is trained using the training dataset for 100 epochs and later validated on the corresponding validation set. After completing 100 epochs, the validation accuracy approximates 0.9. This outcome suggests that Wenxin's understanding of economic development is indeed palpably reflected in its generated urban imagery.

The same methodology was applied to analyze images associated with traditional-modern dimensionality. Following 100 epochs, this detector also recorded a comparable validation accuracy nearing 0.9, thus indicating that Wenxin's perception about whether a city leans towards tradition or modernity is likewise evident in its generated urban landscape visuals. These two detectors are then tested on the test dataset. They are asked to classify images in the test dataset over developed-underdeveloped and traditional-modern dimensions and the results are compared to the ground truth. The confuse matrix is shown in Table 1 and they do a relatively good job in classifying the images in the way that matches Wenxin's understanding.

| Generated  |                | Underdeveloped | Developed |             | Traditional | Modern |
|------------|----------------|----------------|-----------|-------------|-------------|--------|
| Classified | Underdeveloped | 43             | 0         | Traditional | 50          | 3      |
|            | Developed      | 7              | 50        | Modern      | 0           | 47     |

*Table 1: Confusion matrix of the classification results from the test dataset.*

Next, a selection of 25 countries from various continents and at different stages of development are selected for this analysis. Wenxin is asked to generate 100 street view images for cities in each country with unspecified city names. Street view images generated by Wenxin are then passed to the two detectors for classification, thereby ascertaining Wenxin's interpretation of cities in different countries on both developed-underdeveloped and traditional-modern scales.

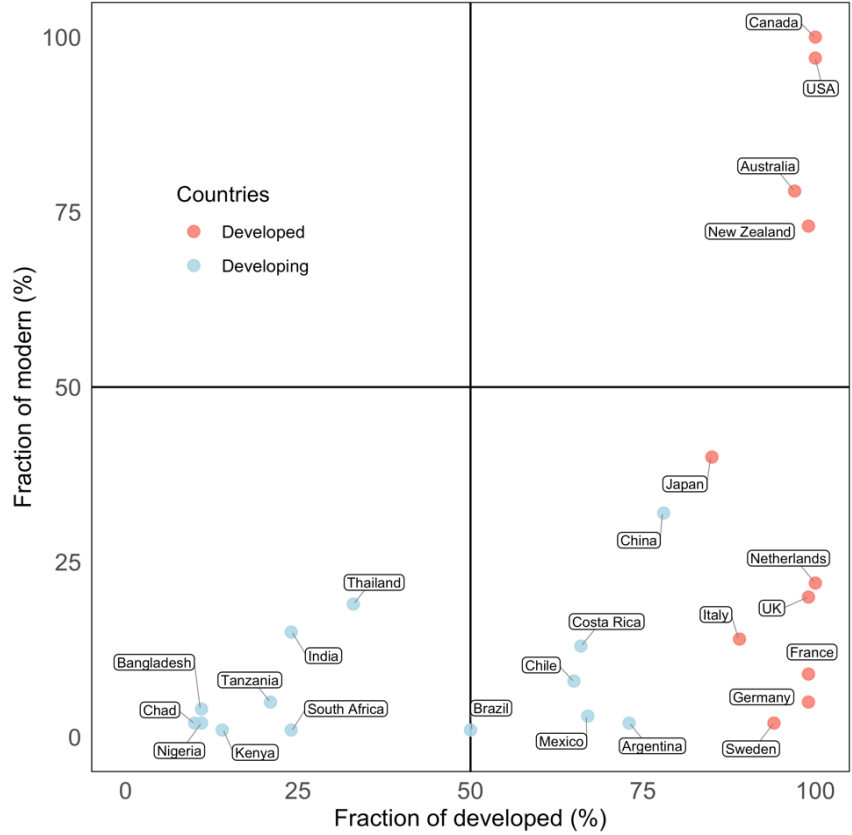

*Figure 2: Cross-plotting of country-level classification results between the two dimensions of generated images for the 25 selected countries.*

**Results**

Classification results from the detectors are depicted in Figure 2. For all selected developed countries, over 80% of the street view images generated by Wenxin are categorized as "economically developed" by the same AI system. It is notable that in developed countries of North American and Oceanian, a significant proportion of images are classified as "modern cities", while those corresponding to developed European countries are predominantly labeled as "traditional cities".

The percentage of generated images identified as "economically developed" ranges between 50% and 75% for Latin American countries; most street views from these regions fall under the "traditional" category within binary classification. The fraction of "economically developed" city imagery associated with China and Japan lies between 75% and 80%, whereas that pertaining to "modern" cities is situated between 30% and 40% for these two Asian countries. For remaining Asian and African nations, both fractions representing 'developed' and 'modern' categories register below the threshold of 50%.

It is important to note that despite the binary nature of classifications on both dimensions, the ResNet-based detectors are capable of providing confidence levels for each image during classification. This capability offers an avenue to examine the distribution of confidence levels across images from different countries, spanning the entire spectrum of underdevelopment-development and traditional-modern dimensions. For reference, results incorporating these confidence levels are displayed in Figure 3.

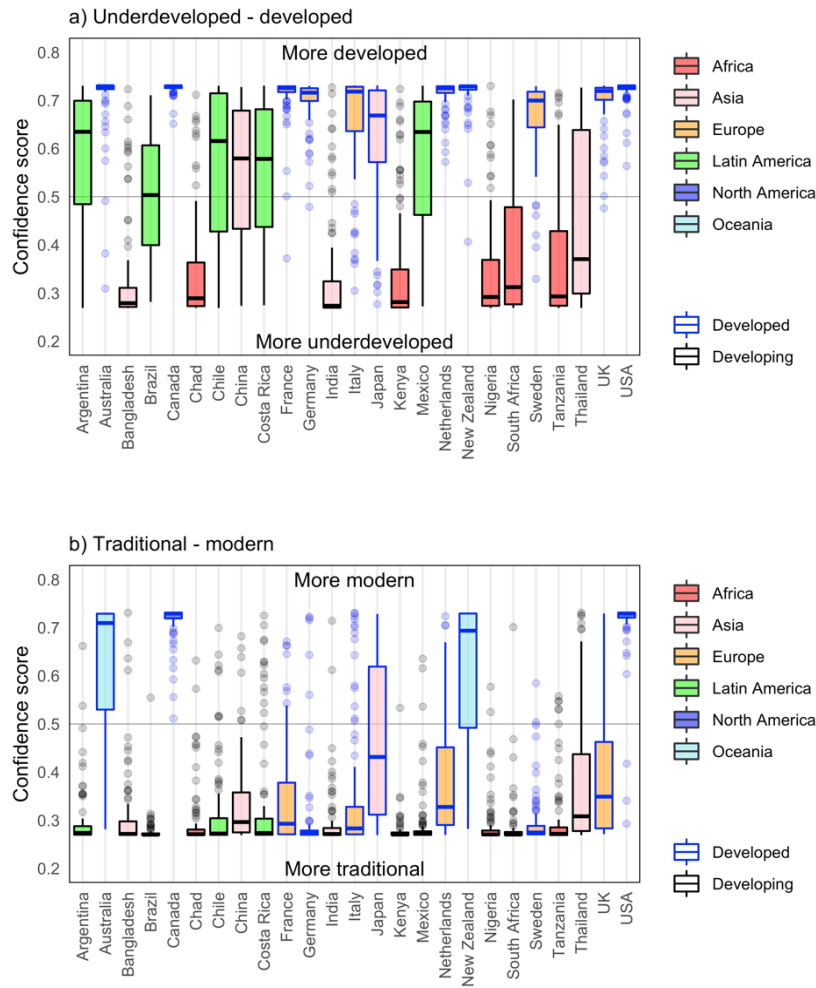

*Figure 3: Distributions of classification confidence levels of generated images on the two dimensions: a) underdeveloped-developed, and b) traditional-modern.*

**Discussion and Conclusion**

It is clearly observable that the classification results of generated street view images from the 25 selected countries cluster along these two dimensions (Figure 2). From Wenxin's perspective, cities in developed North American and Oceanian countries are both economically advanced and characterized by modern-style streetscapes and skyscrapers. Conversely, cities in developed European nations, while economically progressive, retain a greater number of traditional elements. Latin American urban areas in Wenxin's results generally exhibit moderate degrees of economic development alongside a preservation of traditional architecture. Cities in China and Japan display economic advancement with an amalgamation of traditional and contemporary styles within their city landscapes. In contrast, cities in other Asian and African nations are broadly categorized as underdeveloped with a noticeable absence of modern landscapes.

In comparison to real-world scenarios, these results may be influenced by potential stereotypes embedded within the Wenxin model. This AI system appears to summarize economic conditions and city landscapes based on continental-level general regions while also reflecting some national-level characteristics in its generated images. In reality, countries such as the United States possess traditional landscapes across various cities, including those of historical significance like New England. Cities with declining economics are also found in the area of Midwestern United States known as the Rust Belt (e.g. Flynn and Taylor, 1986; Wilson and Heil, 2022). Simultaneously, modern high-rise skylines are prevalent in major cities of developing nations. According to The Skyscraper Center (2023), 14 of the top 20 countries with most skyscrapers are developing countries, and those include China, India, Thailand and Mexico in our selected list. Unfortunately, these diverse urban landscapes are not accurately represented in the street view images produced by Wenxin.

Figure 3 reveals that for the majority of our selected countries, AI-generated images are narrowly confined to a limited confidence-level range against the complete spectrum on both dimensions. However, in actuality, economic development and urban landscape can exhibit significant variations between different cities either across borders or within the same country. This is particularly evident in expansive nations with intricate ethnic and cultural compositions such as China, India, and the United States. Even within singular cities, disparities in economic status and infrastructure development persist among various communities (e.g., You, 2016; Sadeghi and Zanjari, 2017; Lelo et al., 2019; Najib, 2020; Lens, 2022). Consequently, representing all urban landscapes within a country using only a small segment of this spectrum may not yield reliable results. If such AI-generated images were employed for educational purposes or popularizing human geography concepts could potentially reinforce stereotypical impressions about urbanization processes and cityscapes across different countries among students and others seeking knowledge.

The aim of this simple study is to raise awareness about the potential geographic stereotypes that might be embedded in AI-generated images. This doesn't mean we should avoid using generative AI in geography education and outreach activities, but rather it serves as a reminder for users to remain vigilant towards potential biases and stereotypes. At the same time, providers and developers of generative AI should pay more attention to diversity and representation in geographically related knowledge when selecting training data.